\documentclass[aps,pra,preprint, superscriptaddress,showpacs]{revtex4-1}

\usepackage[dvips]{graphicx,color}
\usepackage{times}

\usepackage{bm}        
\usepackage{amssymb}   
\usepackage{amsmath} 
\def\jcp#1#2#3{J.~Chem.~Phys.~{\bf #1},\ #2\ (#3)}

\def\pra#1#2#3{Phys.~Rev.~A~{\bf #1},\ #2\ (#3)}
\def\prl#1#2#3{Phys.~Rev.~Lett.~{\bf #1},\ #2\ (#3)}

\def\jpb#1#2#3{J. Phys. B {\bf #1},\ #2\ (#3)}

\def\k1{k_1}
\def\k2{k_2}
\def\q1{q_1}
\def\q2{q_2}

\def\({\left (}
\def\){\right )}
\def\[{\left [}
\def\]{\right ]}

\newcommand{\beq}{\begin{equation}}
\newcommand{\eeq}{\end{equation}}

\newcommand{\threejm}[6]{ \left(\begin{array}{ccc} #1 & #3 & #5\\
                                             #2 & #4 & #6
                               \end{array}
                         \right)}

\begin{document}
\date{\today}
\flushbottom \draft
\title{Total angular momentum representation for atom-molecule collisions in electric fields}


\author{T. V. Tscherbul}
\affiliation{Harvard-MIT Center for Ultracold Atoms, Cambridge, Massachusetts 02138}
\affiliation{Institute for Theoretical Atomic, Molecular and Optical Physics, Harvard-Smithsonian Center for Astrophysics, Cambridge, Massachusetts 02138}\email[]{tshcherb@cfa.harvard.edu}

\begin{abstract}
It is shown that the atom-molecule collision problem in the presence of an external electric field can be solved using the total angular momentum representation in the body-fixed coordinated frame, leading to a computationally efficient method for {\it ab initio} modeling of low-temperature scattering phenomena.  
 Our calculations demonstrate rapid convergence of the cross sections for vibrational and Stark relaxation in He-CaD collisions with the number of total angular momentum states in the basis set, leading to a 5-100 fold increase in computational efficiency over the previously used methods based on the fully uncoupled space-fixed representation. These results open up the possibility of carrying out numerically converged quantum scattering calculations on a wide array of atom-molecule collisions and chemical reactions in the presence of electric fields.

\end{abstract}

\maketitle

\clearpage
\newpage

\section{Introduction}

Recent experimental and theoretical studies have shown that external electromagnetic fields can be used as a powerful tool to manipulate molecular collisions and chemical reactivity at low temperatures \cite{Book,NJP,Schnell,Softley,Dulieu,JohnCPC,Roman08,BalaJPB04,prl06,jcp07,jcp06,Sergey09,jcp08,MeyerBohn,KRb1,KRb2,KRb_theory}. Examples include resonant control of atom-molecule collisions and chemical reactions in ultracold molecular gases \cite{BalaJPB04,prl06,jcp06,Sergey09}, electric field control of nascent product state distributions \cite{jcp08,MeyerBohn}, and off-resonant laser field control of motional degrees of freedom \cite{KRb_theory,KRb2}. These pioneering studies demonstrate that future progress in the field of cold molecules -- in particular, the ability to create large, dense, and stable ensembles of chemically diverse molecular species -- will depend to a large extent on our understanding of their collisional properties \cite{Book,NJP,Schnell,Softley,Dulieu,JohnCPC,Roman08}. 


Theoretical modeling of molecular collision experiments performed at temperatures below 1~K requires quantum scattering calculations based on multidimensional potential energy surfaces (PESs) of unprecedented accuracy, which generally remain beyond the capabilities of modern {\it ab initio} methods. A way out of this difficulty is to adjust the interaction PESs based on experimental measurements of collision observables such as trap loss rates \cite{CaH,RbCs,He-OH,KRb1,KRb2,FeshbachChemistry,NH,NHexp,N-NH,OH-ND3,Rb-NH3}.   The crucial link between intermolecular PESs and laboratory observations is provided by quantum scattering calculations, which yield collisional properties of molecules exactly for a given PES. Because of the need to incorporate symmetry breaking effects arising from the presence of external fields \cite{Roman08}, such calculations are more challenging  than their field-free counterparts. In particular, the total angular momentum of the collision pair is no longer conserved in the presence of external fields, invalidating the standard approaches of molecular collision theory based on the total angular momentum representation \cite{AD,Lester}.


A theoretical formalism for quantum scattering calculations of molecular collisions in external fields was developed by Volpi and Bohn and by Krems and Dalgarno \cite{VolpiBohn,Roman04}.  The formalism is based on the fully uncoupled space-fixed representation, in which the wavefunction of the collision complex is expanded in direct products of rotational basis functions and spherical harmonics describing the orbital motion of the collision partners in a space-fixed (SF) coordinate frame \cite{VolpiBohn,Roman04}. Several groups have used this representation to study the effects of external electric, magnetic, and microwave fields on atom-molecule \cite{prl06,NH,jcp06,Jeremy07,HeCH2,Thierry,Chinese} and molecule-molecule  \cite{njp09,JesusO2,Liesbeth} collisions. These studies have shown that the fully uncoupled SF formalism meets with serious difficulties when applied to collision problems characterized by strongly anisotropic interactions \cite{njp09,jcp08,Li-NH}. More specifically, the interaction anisotropy strongly couples different rotational and partial wave basis states, leading to very large systems of coupled-channel equations that are beyond the capability of present-day computational resources. As most atom-molecule and molecule-molecule interactions are strongly anisotropic, this difficulty has precluded converged calculations on many interesting collision systems, including Li + HF  $\leftrightarrow$  LiF + H \cite{jcp08}, Rb + ND$_3$ \cite{Rb-NH3}, Li + NH \cite{Li-NH}, and NH + NH  \cite{Liesbeth}. 



We have recently developed an alternative approach to atom-molecule and molecule-molecule scattering  in a magnetic field based on the total angular momentum representation \cite{jcp10}. The total angular momentum of the collision complex is approximately conserved even in the presence of external fields; thus, using basis functions with well-defined total angular momentum allows for a substantial reduction in the number of scattering channels \cite{jcp10}. This advantage allowed us to obtain numerally converged scattering cross sections for strongly anisotropic atom-molecule \cite{Li-CaH} and molecule-molecule \cite{Yura} collisions in the presence of a magnetic field. 
Magnetic fields interact with the electron spin of the molecule, which can be weakly coupled to the intermolecular axis and often plays a spectator role during the collision.  As a result, while  an applied magnetic field shifts the energies of the colliding molecules and may lead to the appearance of scattering resonances, it hardly affects the mechanism of collision-induced energy transfer. In contrast, electric fields break the inversion symmetry of the collision problem and alter the selection rules for parity-changing transitions, leading to more dramatic changes in  collision mechanisms. Examples include electric field-induced molecular states \cite{AvdeenkovBohn},  dipolar resonances  \cite{Ticknor}, enhancement and suppression of spin relaxation in $^2\Sigma$ and $^2\Pi$ molecules \cite{prl06,FaradayDiscuss}, and stimulated chemical reactions  \cite{KRb1,KRb_theory}. 


The purpose of this article  is to extend the approach developed in Ref. \cite{jcp10} to describe atom-molecule collisions in electric fields. 
In Sec. II, we formulate the collision problem in the total angular momentum representation and outline the procedure of evaluating atom-molecule collision cross sections. We then apply our formulation to calculate the cross sections for Stark relaxation (Sec. IIIA) and vibrational relaxation (Sec. IIIB) in $^3$He-CaD collisions in the presence of an electric field. Our results agree well with benchmark calculations based on the  fully uncoupled SF representation, demonstrating the validity and efficiency of our approach. These findings lead us to conclude that numerical algorithms based on the total angular momentum representation are a powerful way of carrying out quantum scattering calculations in the presence of electric fields. Sec. IV presents a brief summary of main results and outlines future research directions opened up by this work.

\section{Theory}




A non-reactive collision of a diatomic molecule (BC) with a structureless atom (A) in the presence of a dc electric field is described by the Hamiltonian (in atomic units) \cite{jcp10}
\begin{equation}\label{H}
\hat{H} = -\frac{1}{2\mu R}\frac{\partial^2}{\partial R^2} R + \frac{\hat{\ell}^2}{2\mu R^2} + V(\mathbf{R},\mathbf{r}) + \hat{H}_\text{as},
\end{equation}
where $\mathbf{R}$ is the atom-molecule separation vector, $\mathbf{r}=r\hat{r}$ defines the length and the orientation of the internuclear axis (BC) in the SF frame, $\hat{\ell}$ is the orbital angular momentum for the collision,  $V(\mathbf{R},\mathbf{r})$ is the atom-molecule interaction potential, and $\mu$ is the A-BC reduced mass,  The asymptotic Hamiltonian $\hat{H}_\text{mol}$ describes the rovibrational structure of the diatomic molecule and its interaction with an electric field of strength $E$  oriented along the SF quantization axis $Z$
\begin{equation}\label{Has}
\hat{H}_\text{as} = -\frac{1}{2mr} \frac{d^2}{dr^2}r + \frac{\hat{\jmath}^2}{2m r^2} + V(r) -  Ed\cos\theta_r
\end{equation}
where $\hat{\jmath}$ is the rotational angular momentum, $d$ is the permanent electric dipole moment of the molecule with mass $m$, $V(r)$ is the intramolecular interaction potential \cite{BalaHeCaH_paper1}, and $\theta_r$ is the polar angle of the internuclear axis ($\hat{r}$) in the SF frame \cite{prl06,jcp06}.

 The orbital angular momentum $\hat{\ell}^2$ in Eq. (\ref{H}) can be expressed via the total angular momentum of the collision complex $\hat{J}$ in the body-fixed (BF) coordinate frame as \cite{jcp10,Lester}
\begin{equation}\label{l2}
\hat{\ell}^2  = (\hat{J} - \hat{\jmath})^2 = \hat{J}^2 + \hat{\jmath}^2 - \hat{J}_+\hat{\jmath}_- - \hat{J}_-\hat{\jmath}_+ - 2\hat{J}_z\hat{\jmath}_z,
\end{equation}
where $\hat{J}_\pm$ and $\hat{\jmath}_\pm$ are the BF raising and lowering operators (note that $\hat{J}_\pm$ satisfy anomalous commutation relations \cite{Zare}). The BF $z$-axis coincides with the vector $\mathbf{R}$ and the $y$-axis is perpendicular to the collision plane.

As in our previous work \cite{jcp10}, we expand the wave function of the collision complex in direct products of BF basis functions \cite{Lester,jcp10}
\begin{equation} \label{BFexpansion}
\Psi = \frac{1}{R}\sum_{J}\sum_{v,\, j,\,k} F^M_{Jvjk}(R) |vjk\rangle |JMk\rangle, 
\end{equation}
where $k$ is the BF the projection of $J$ and $j$, and $M$ is the SF projection of $J$. In Eq. (\ref{BFexpansion}), $|JMk\rangle=\sqrt{(2J+1)/8\pi^2}D^{J*}_{Mk}(\Omega_E)$ are the  symmetric top eigenfunctions, $D(\Omega_E)$ are the Wigner $D$-functions, and $\Omega_E$ are the Euler angles which specify the orientation of BF axes in the SF frame. The functions $|vjk\rangle=r^{-1}\chi_{vj}(r)\sqrt{2\pi}Y_{jk}(\theta,0)$ describe the rovibrational motion of the diatomic molecule in the BF frame. The rovibrational functions $\chi_{vj}(r)$ satisfy the Schr{\"o}dinger equation
\begin{equation}\label{chi}
\left[-\frac{1}{2m} \frac{d^2}{dr^2} + \frac{j(j+1)}{2 mr^2} + V(r)\right] \chi_{vj}(r) = \epsilon_{vj}\chi_{vj}(r)
\end{equation}
where $\epsilon_{vj}$ is the rovibrational energy of the molecule in the absence of an electric field \cite{note}.

The radial expansion coefficients $F^M_{Jvjk}(R)$ satisfy a system of coupled-channel (CC) equations 
\begin{align}\label{CC}\notag
\left[ \frac{d^2}{dR^2} +2\mu E_\text{tot}\right] F^M_{Jvjk}(R)=2\mu \sum_{J',\,v',j',k'} \langle  JMk | \langle vjk | V(R,r,\theta) &+ \frac{1}{2\mu R^2} (\hat{J} - \hat{\jmath})^2 \\&+\hat{H}_\text{as} | J'Mk'\rangle |v'j'k'\rangle F^M_{J'v'j'k'}(R),
\end{align}
where $E_\text{tot}$ is the total energy. The matrix elements of the interaction potential and of $\hat{\ell}^2$ can be evaluated as described in Refs. \cite{Lester,jcp10}. In the absence of an electric field, the asymptotic Hamiltonian (\ref{Has}) has only diagonal matrix elements 
\begin{equation} \label{me_rot}
\langle  JMk | \langle vjk | \hat{H}_\text{as}| J'M'k'\rangle |v'j'k'\rangle = \delta_{JJ'}\delta_{MM'}\delta_{vv'}\delta_{jj'}\delta_{kk'} \epsilon_{vj} \,\,\, (E=0).
\end{equation}

In order to evaluate the matrix elements of the molecule-field interaction in the BF basis, we transform the $Z$-component of vector $\hat{r}$ to the BF frame  \cite{Zare}
\begin{equation}\label{transformation}
\cos\theta_r = \left(\frac{4\pi}{3}\right)^{1/2} Y_{10} (\theta_r,\phi_r) = \left(\frac{4\pi}{3}\right)^{1/2}  \sum_q D^{1*}_{0q}(\Omega_E) Y_{1q} (\theta,\phi).
\end{equation}
The expression on the right-hand contains spherical harmonics of BF angles ($\theta,\phi$) and Wigner $D$-functions of Euler angles (note that $\theta$ is the Jacobi angle between $\mathbf{R}$ and $\mathbf{r}$). Making use of standard expressions for angular integrals involving three spherical harmonics \cite{Zare}, and neglecting the $r$ dependence of $d$ (which is a good approximation for low vibrational states and weak electric fields \cite{Rosario}) we obtain for the molecule-field interaction matrix element 
\begin{multline}\label{me_ext}
\langle JMk | \langle vjk | -Ed\cos\theta_r | J'M'k' \rangle |v'j'k'\rangle =
-Ed \delta_{MM'} \delta_{vv'} [(2J+1)(2J'+1)(2j+1)(2j'+1)]^{1/2} \\ \times   (-)^{M+k-k'}\sum_q  (-)^q
 \threejm{J}{M}{1}{0}{J'}{-M} \threejm{j}{0}{1}{0}{j'}{0} 
 \threejm{J}{k}{1}{-q}{J'}{-k'} \threejm{j}{-k}{1}{q}{j'}{k'}.
\end{multline}
This expression shows that the interaction with electric fields couples basis functions of different $J$. It is because of this coupling that the collision problem can no longer be factorized by symmetry into smaller $J$-subproblems \cite{Lester}. 
It follows from Eq. (\ref{me_ext}) that (i) the external field couplings vanish unless $J-J' = \pm 1$, and (ii) electric fields couple basis functions of different $k$,  leading to a field-induced analog of the Coriolis interaction. Unlike the standard Coriolis interaction, however, the interaction with external electric fields couples different $k$-states in {\it different} $J$-blocks (assuming $M=0$). 



The standard asymptotic analysis of the radial solutions to CC equations (\ref{CC}) at large $R$ gives the $S$-matrix elements and scattering observables. The analysis proceeds in two steps.  First, the BF wavefunction is transformed to the SF representation using the eigenvectors of the operator $\hat{\ell}^2$ \cite{jcp10, ABC,Launay}.  Next,  the wavefunction is transformed to the basis in which $\hat{H}_\text{as}$ is diagonal using the eigenvectors of the asymptotic Hamiltonian (\ref{Has}) in the SF representation. The eigenvalues of $\hat{H}_\text{as}$ define the scattering channels $|\gamma \ell\rangle$ and threshold energies $\epsilon_\gamma$ in the presence of an electric field. 

Matching the transformed solutions to the asymptotic form \cite{jcp10}
\begin{equation}\label{BoundaryConditions}
F^M_{\gamma \ell}(R) \to  \delta_{\gamma\gamma'}\delta_{\ell\ell'} e^{-i(k_\gamma R - \ell \pi/2)} - \left(\frac{k_{\gamma}}{k_{\gamma'}}\right)^{1/2} S^M_{\gamma \ell; \gamma'\ell'}e^{i(k_{\gamma'} R - \ell' \pi/2)}
\end{equation}
yields the $S$-matrix elements describing collision-induced  transitions between the channels $\gamma$ and $\gamma'$ with wavevectors $k^2_\gamma = 2\mu(E_\text{tot}-\epsilon_\gamma)=2\mu E_C$, where $E_C$ is the collision energy. The integral cross sections can be evaluated from the $S$-matrix elements as \cite{Roman04, jcp10}
\begin{equation}
\sigma_{\gamma \to\gamma'} = \frac{\pi}{k_\gamma^2} \sum_M \sum_{\ell,\,\ell'} |\delta_{\ell\ell'}\delta_{\gamma\gamma'} - S^M_{\gamma \ell; \gamma\ell'} |^2.
\end{equation}


For the He-CaD interaction, we used a three-dimensional {\it ab initio} potential energy surface developed by Balakrishnan {\it et al.} \cite{BalaHeCaH_paper1,BalaHeCaH}, which explicitly includes the $r$ dependence of the interaction energy. The rovibrational eigenfunctions $\chi_{vj}(r)$ were evaluated by solving the one-dimensional Schr{\"o}dinger equation (\ref{chi}) using a  discrete variable representation (DVR) method \cite{ColbertMiller}. The matrix elements of the He-CaD interaction in Eq. (\ref{CC}) were obtained by expanding the PES in Legendre polynomials with $\lambda_\text{max}=12$ and evaluating the integrals over spherical harmonics analytically to yield \cite{Lester,jcp10}
\begin{multline}
 \langle  JMk | \langle vjk | V(R,r,\theta) | J'Mk'\rangle |v'j'k'\rangle = \delta_{JJ'}\delta_{kk'} [(2j+1)(2j'+1)]^{1/2} \\ \times
  \sum_{\lambda=0}^{\lambda_\text{max}} \langle \chi_{vj}(r) | V_\lambda (R,r) |\chi_{v'j'}(r)\rangle \threejm{j}{-k}{\lambda}{0}{j'}{k'} \threejm{j}{0}{\lambda}{0}{j'}{0}
\end{multline}
The radial coefficients $V_\lambda (R,r)$ were evaluated using a 24-point Gauss-Legendre quadrature in $\theta$. The $r$ integrals were computed with 30 Gauss-Legendre quadrature points in $r \in [2.5, 5.6]$ $a_0$.

The CC equations (\ref{CC}) were solved using the log-derivative method \cite{David86} on a grid of $R$ between 2 and 100~$a_0$ with a grid step of 0.1 $a_0$. The BF basis set used in  Stark relaxation calculations (Sec. IIIA) included 10 rotational states ($j_\text{max}=9$); the basis set used in vibrational relaxation calculations included 10 rotational states in  $v=0$ and $v=1$ vibrational manifolds of CaD (see Sec. IIIB). The cross sections for Stark relaxation were converged to $<$10\%. 

For classification purposes, the eigenvalues of the asymptotic Hamiltonian are assigned physical quantum numbers appropriate to a polar diatomic molecule in an electric field: $v$, $j$, and $m$ (the SF projection of $j$). In this work, we are interested in low-to-moderate field strengths, where the interaction with electric field is small compared to the splitting between the ground and the first excited rotational levels. We can therefore keep using $j$ to denote the rotational manifold and $m$ to distinguish the Stark states within the manifold, even though $j$ is not a good quantum number in an electric field. The assignment procedure works as follows. All eigenvalues of the asymptotic Hamiltonian which are close in energy to a particular Stark state $|vjm\rangle$  (that is, $|\epsilon_\gamma-\epsilon_{vjm}|<\Delta$) are assigned the quantum numbers $v,j,m$. The eigenvalues that do not meet this condition are excluded from consideration. In this work, we set $\Delta=0.1$ cm$^{-1}$, however, test calculations show that the results are not sensitive to the choice of $\Delta$ as long as $E_C<\Delta$. If this condition is not met,  problems may arise with distinguishing between elastic and inelastic channels (see Sec. IIIB). 



\section{Results and discussion}

In this section, we first consider the eigenstates of the asymptotic Hamiltonian that define the scattering channels in the presence of an electric field (Sec. IIIA). In order to test the performance of our approach, we compare the cross sections calculated using the BF total angular momentum representation with benchmark calculations based on the fully uncoupled SF representation (Secs. IIIB and C). 

\subsection{Asymptotic states}

\begin{figure}[t]
	\centering
     \includegraphics[width=0.60\textwidth, trim = 0 0 0 0]{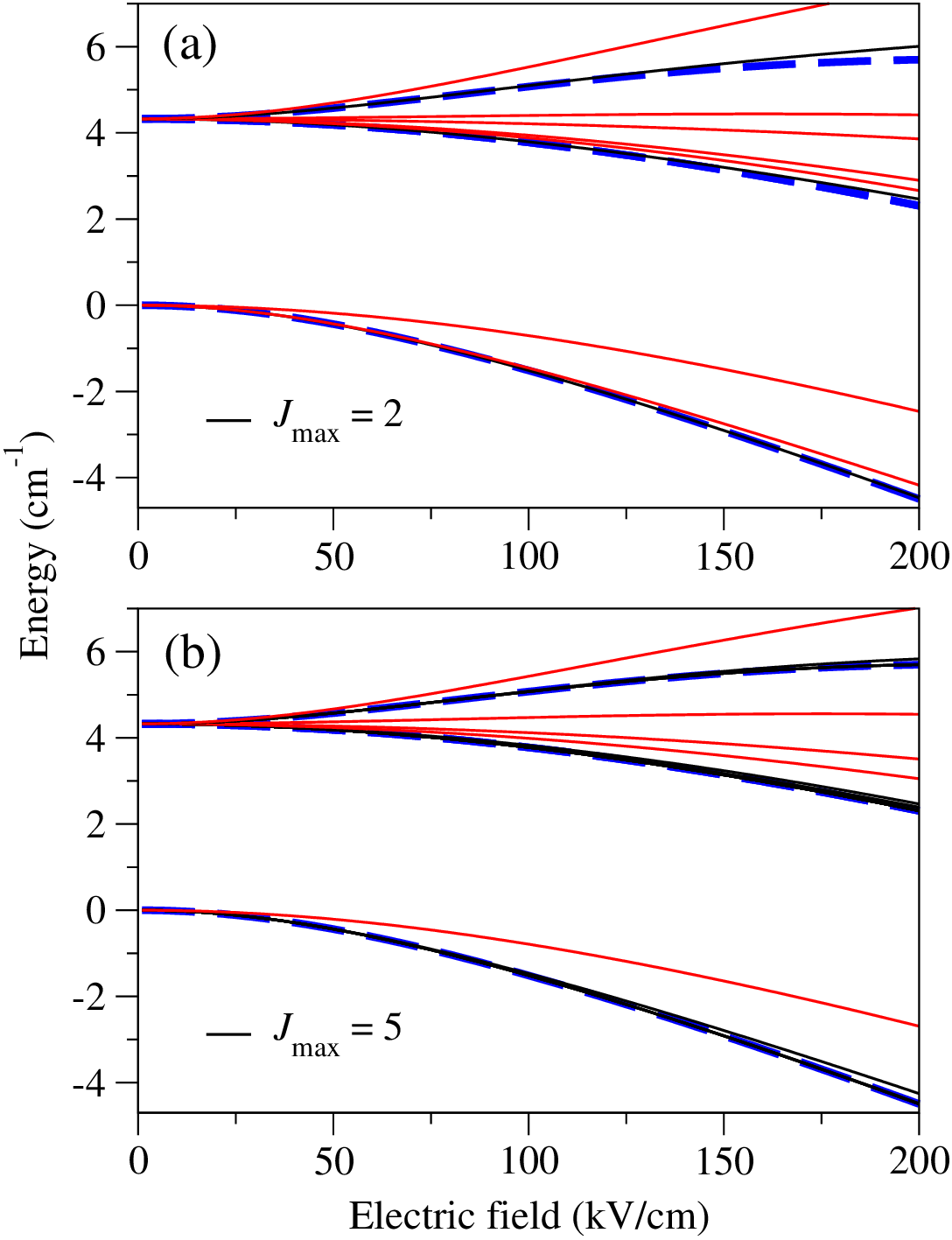}
	\renewcommand{\figurename}{Fig.}
	\caption{Stark levels of CaD (bold dashed lines) and eigenvalues of the asymptotic Hamiltonian (full lines) as functions of the applied electric field. Upper panel: calculation with $J_\text{max}=2$, lower panel: calculation with $J_\text{max}=5$. Unphysical states are shown by red lines.  Both calculations are for $j_\text{max}=9$. }
	\label{fig:stark}
\end{figure}

Figure \ref{fig:stark} shows the eigenvalues of the asymptotic Hamiltonian (\ref{Has}) for the ground vibrational state of CaD as functions of the applied electric field. The number of total $J$-states is given by $N_J=J_\text{max}+1$, where $J_\text{max}$ is the largest value of $J$ included in the basis set. The eigenvalues obtained for $J_\text{max}=2$ and 5 are shown in the upper and lower panels, respectively. The results clearly show that $\hat{H}_\text{as}$ expressed in the total angular momentum basis  has  eigenvalues that do not correspond to the physical Stark states of the diatomic molecule. This situation is similar to that encountered in the case of magnetic fields, and following the terminology introduced in \cite{jcp10}, we will refer to these states as "unphysical". From Fig. \ref{fig:stark}, we observe that the number of unphysical Stark states increases with the number of $J$-blocks in the basis set. In addition, the energies of the unphysical states become closer to the true Stark energies as $J_\text{max}$ increases.

As pointed out before \cite{jcp10}, the origin of the unphysical states shown in Fig. \ref{fig:stark} can be attributed to the basis set truncation procedure. The total $J$ basis is truncated by restricting the number of $J$-blocks ($N_J=J_\text{max}+1)$. However, as follows from Eq. (\ref{me_ext}), electric fields couple basis states in block $J$ to those in block $J+1$. When the Hamiltonian matrix is truncated, these couplings are left out, resulting in the appearance of unphysical eigenvalues and eigenvectors. In Ref. \cite{jcp10} it was shown that the eigenvectors of unphysical Zeeman states are dominated by the largest value of $J$ included in the basis set. As a result, the presence of unphysical states has no influence on low-temperature collisions in magnetic fields \cite{jcp10}. 

In order to elucidate the properties of unphysical  states, we consider the matrix of the asymptotic Hamiltonian (\ref{Has}) in the BF basis. In the weak-field limit $|Ed|/B_e\ll 1$, we can consider only the coupling between the ground and the first excited rotational states in the $v=0$ manifold (the $v$ index will be omitted for the rest of this section). Arranging the $|JMk\rangle |jk\rangle$ functions in the following sequence: $|000\rangle |00\rangle$, $|100\rangle |1-1\rangle$,   $|100\rangle |10\rangle$,  $|100\rangle |11\rangle$, $|000\rangle |10\rangle$, $|100\rangle |00\rangle$,  we obtain the matrix of the asymptotic Hamiltonian 
\begin{equation}\label{Matrix}
\left(\begin{array}{cc} H_1 & 0 \\
                               0    & H_2 \\            
                               \end{array}\right),
\end{equation}  
with
\begin{equation}\label{H1} H_1 =
\left(\begin{array}{cccc}  0  &    -\frac{1}{3}Ed  & -\frac{1}{3}Ed & -\frac{1}{3}Ed  \\
                              -\frac{1}{3}Ed&   2B_e         &  0      &  0  \\
                               -\frac{1}{3}Ed &   0  &   2B_e      &  0  \\
                               -\frac{1}{3}Ed &   0  &  0     &  2B_e                
                               \end{array}\right)
\end{equation}   
and
\begin{equation}\label{H1} H_2 =
\left(\begin{array}{cc}  2B_e  &   -\frac{1}{3}Ed   \\
                             -\frac{1}{3}Ed &   0              
                               \end{array}\right),
\end{equation}  
Diagonalization of $H_1$ yields
\begin{equation}
\lambda_{1,2} = B_e \pm \sqrt{B_e^2 +\frac{1}{3}(Ed)^2}, \quad \lambda_{3,4} = 2B_e.
\end{equation}
These energies are the same as those of a polar $^1\Sigma$ molecule in a dc electric field \cite{BohnChapter,jcp07}. 
The eigenvalues of $H_2$ 
\begin{equation}\label{lambda56}
\lambda_{\pm} = B_e \pm \sqrt{B_e^2 +\frac{1}{9}(Ed)^2}.
\end{equation}
correspond to unphysical Stark states. The eigenvectors of the unphysical states are given by
\begin{equation}\label{lambda_pm}
|\lambda_\pm\rangle = \frac{-\frac{1}{3}Ed}{D_\pm} |000\rangle |00\rangle + \frac{B_e\mp \sqrt{B_e^2+\frac{1}{9}(Ed)^2}}{D_\pm} |100\rangle |00\rangle
\end{equation}
where $D_\pm^2 =  (Ed)^2 + \left[ B_e - \sqrt{B_e^2 \mp (Ed)^2} \right]^2$. Eq. (\ref{lambda_pm}) illustrates that the field-induced mixing between different $J$-states is proportional to the magnitude of the electric field. Thus, we expect that the coupling between the different $J$-blocks will become stronger with increasing field, making it necessary to include more $J$-blocks in the basis set to obtain converged results even at ultralow collision energies (see Sec. III). By contrast, the eigenvectors of unphysical Zeeman states are, to a first approximation, independent of the field strength \cite{jcp10}, and so are the convergence properties of scattering observables.


Finally, we note that neglecting the electric-field-induced coupling within the $H_2$ block leads to the disappearance of unphysical Stark shifts (\ref{lambda_pm}). This observation suggests a  way to eliminate the unphysical states from scattering calculations. Preliminary results obtained with a restricted basis set ($j_\text{max}=1$, $J_\text{max}=1$) indicate that neglecting the off-diagonal elements of $H_2$ does provide accurate results for both the elastic and inelastic He-CaD scattering. It remains to be seen whether or not the procedure can be generalized to  larger  rotational basis sets.

\subsection{Stark relaxation in He-CaD$(v=0,j=1,m_j=0)$ collisions}

\begin{figure}[t]
	\centering
     \includegraphics[width=0.60\textwidth, trim = 0 0 0 0]{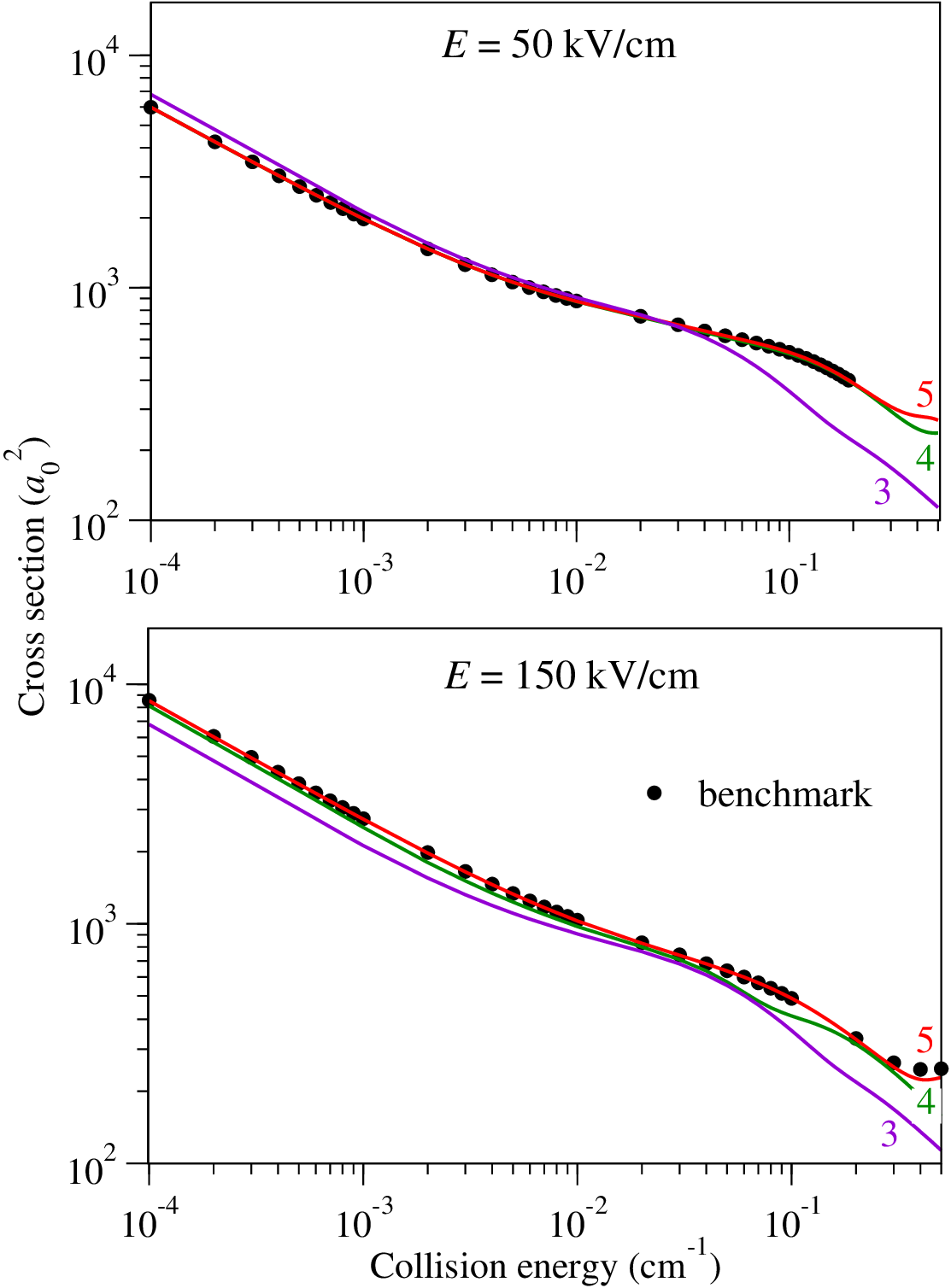}
	\renewcommand{\figurename}{Fig.}
	\caption{Cross sections for Stark relaxation in He-CaD($v=0,j=1, m=0$) collisions as functions of collision energy. The curves are labeled by the maximum value of $J$ included in the basis set ($J_\text{max}$),  see text for details. The electric field is 50 kV/cm (upper panel) and 150 kV/cm (lower panel). The calculations are performed for the total angular momentum projection $M=0$. Circles -- benchmark results obtained using the fully uncoupled SF representation.}
	\label{fig:rot}
\end{figure}

Figure \ref{fig:rot} shows the cross sections for Stark relaxation in $^3$He-CaD$(v=0,j=1,m_j=0)$ collisions calculated using the BF total angular momentum representation. The inelastic cross sections are summed over all final Stark states of CaD and displayed as functions of collision energy for $M=0$. At very low collision energies (in the Wigner s-wave limit) the cross sections scale as $1/\sqrt{E_C}$ \cite{Book,Roman08}. At higher collision energies, the cross sections display broad oscillations due to the presence of scattering resonances \cite{jcp06,jcp08}.

At an electric field of 50 kV/cm, the BF results obtained with $J_\text{max}= 5$ are in excellent agreement with the benchmark calculations over the entire range of collision energies from $10^{-4}$ cm$^{-1}$ to 1 cm$^{-1}$. The agreement for $J_\text{max}=4$ is also good at $E_C>0.1$ cm$^{-1}$. The deviations observed above this collision energy occur because the  number of total angular momentum states in the basis is not sufficient to adequately describe scattering resonances in the entrance and/or exit collision channels. This is analogous to the lack of convergence at high collision energies observed in our previous calculations of atom-molecule collisions in magnetic fields \cite{jcp10}. The cross sections obtained with  $J_\text{max}=3$ are off by $\sim$50~\% even in the $s$-wave regime, which indicates that the external field coupling between the  $J=3$ and $J=4$ blocks can no longer be neglected.

In order to test the performance of our algorithm at higher electric fields, we display in the lower panel of Fig. \ref{fig:rot} the cross sections calculated for $E=150$ kV/cm for different values of $J_\text{max}$. While $J_\text{max}=4$ cross sections display a similar energy dependence as the benchmark results, quantitative agreement requires extension of the basis set to $J_\text{max}=5$. We conclude that it is necessary to include more $J$-states in the basis set to achieve convergence at higher electric fields. 

 
 As shown in the previous section, the properties of  unphysical Stark states depend on the magnitude of the electric field. At higher electric fields the scattering wavefunction contains contributions from higher $J$-blocks, making it necessary to  increase $J_\text{max}$ to obtain converged results even at ultralow collision energies, as illustrated by the results plotted in Figs. \ref{fig:rot} and \ref{fig:vib}. By contrast, converged results for ultracold atom-molecule collisions in magnetic fields can typically be obtained with just two $J$-blocks \cite{jcp10}.




\subsection{Vibrational relaxation: He-CaD$(v=1,j=0,m_j=0)$ collisions}

\begin{figure}[t]
	\centering
     \includegraphics[width=0.60\textwidth, trim = 0 0 0 0]{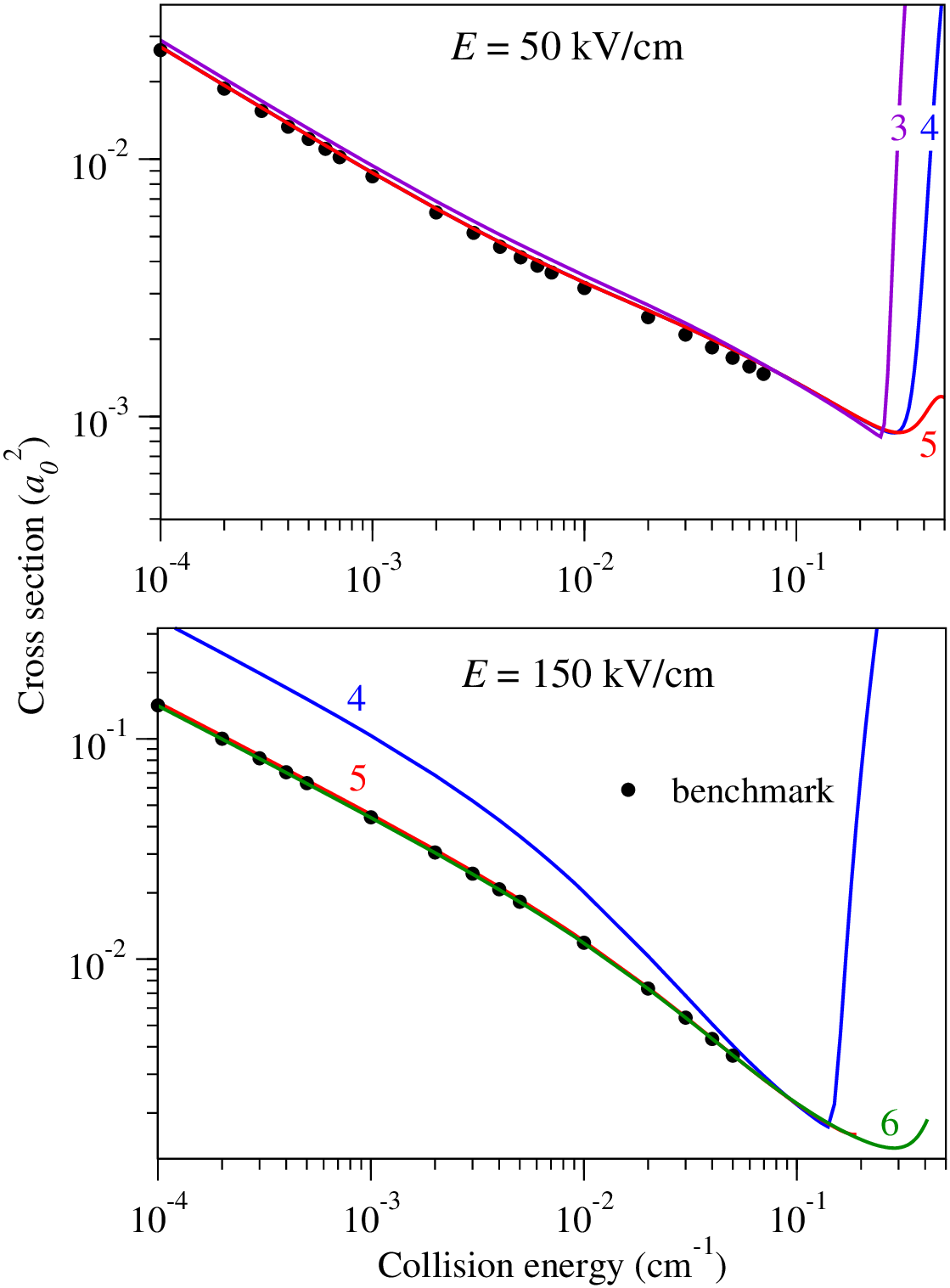}
	\renewcommand{\figurename}{Fig.}
	\caption{Cross sections for vibrational relaxation in He-CaD($v=1,j=0, m=0$) collisions summed over all final Stark states of CaD as functions of collision energy. The curves are labeled by the maximum value of $J$ included in the basis set ($J_\text{max}$). The electric field is 50 kV/cm (upper panel) and 150 kV/cm (lower panel). The calculations are performed for the total angular momentum projection $M=0$.}
	\label{fig:vib}
\end{figure}

In Fig. \ref{fig:vib}, we compare the cross sections for vibrational relaxation in He-CaD($v=1,j=0,m_j=0$) collisions calculated using the BF approach with benchmark SF calculations. The cross sections are summed over all final rotational states of CaD as plotted as functions of collision energy for different $J_\text{max}$.
 Balakrishnan {\it et al.} considered vibrational relaxation in $^3$He-CaH($v=1,j=0$) collisions in the absence of external fields and found it necessary to include 20 rotational states in the $v=0$ and $v=1$ vibrational manifolds to achieve numerical convergence \cite{BalaHeCaH}. The first excited vibrational state of CaD lies 908.3 cm$^{-1}$ above the ground state, and the rotational constant of CaD is 2.16 cm$^{-1}$. In order to properly describe quasiresonant energy transfer important at low temperatures \cite{Bala98}, it would thus be necessary to include at least 20 rotational states of CaD in each vibrational manifold. A fully uncoupled SF basis with $v_\text{max}=1$, $j_\text{max}=20$, and $l_\text{max}={20}$ contains 12362 channels. In order to avoid solving large numbers of CC equations, we opted to use a restricted SF basis set with $v_\text{max}=1$, $j_\text{max}=9$, and $l_\text{max}=9$ to generate benchmark results, which should be adequate for testing purposes provided the same convergence parameters $v_\text{max}$ and $j_\text{max}$ are used in BF and SF calculations. We emphasize, however, that these benchmark cross sections are not physically meaningful (e.g., they may not exhibit the quasi-resonance behavior characteristic of vibrational relaxation at low temperatures \cite{BalaHeCaH,Bala98}). 
 

From Fig. \ref{fig:vib} we observe that the BF cross sections obtained for a relatively weak electric field ($E=50$ kV/cm) are in good agreement with benchmark calculations already at $J_\text{max}=3$. Table I demonstrates that a $J_\text{max}=3$  calculation includes only 280 scattering channels, while the same calculation performed using the fully uncoupled SF representation requires as many as 1380 channels. The use of the BF total angular momentum representation thus allows us to reduce the number of scattering channels by a factor of 4. The computational cost of solving CC equations scales as $N^3$ with the number of scattering channels \cite{David86}, so the BF total angular momentum representation is more than 100-fold more computationally efficient than the fully uncoupled SF representation \cite{VolpiBohn,Roman04}.

At $E=150$ kV/cm, quantitatively accurate results are obtained with $J_\text{max}\ge 5$, while $J_\text{max}=4$ calculations overestimate the benchmark result by a factor of $\sim$3.  
Comparison of Figs. \ref{fig:rot} and \ref{fig:vib} suggests that vibrational relaxation cross sections converge more slowly with $J_\text{max}$ than those for Stark relaxation. 
The gain in computational efficiency ($\sim$10-fold) is therefore not as dramatic as observed for $E=50$ kV/cm. Note that the BF inelastic cross sections show an unphysical jump at a collision  energy of $\sim$0.14 cm$^{-1}$. This jump occurs because of the ambiguity of the procedure used to assign quantum numbers to unphysical states. As pointed out in Sec. II, the eigenvalues of the asymptotic Hamiltonian with energies $|\epsilon_\gamma-\epsilon_{vjm}|<\Delta$ are assigned physical quantum numbers $v,j$, and $m$, where we have chosen $\Delta=0.1$ cm$^{-1}$. While this procedure works well as long as the collision energy is small compared to $\Delta$, collision-induced transitions between unphysical states make it difficult to distinguish between elastic and inelastic channels when this condition is not met. This technical difficulty can be eliminated by increasing $\Delta$ or switching to an unphysical states-free representation (see~Sec.~II).


\section{Summary}

We have presented an efficient theoretical approach to solving the atom-molecule collision problem in the presence of an electric field. Unlike previous theoretical work based on the fully uncoupled space-fixed representation \cite{Roman04,VolpiBohn}, our approach makes explicit use of the total angular momentum ($J$) representation in the body-fixed coordinate frame, in which the atom-molecule Hamiltonian has a block-diagonal form in the absence of external fields. The different $J$ blocks are coupled only by the molecule-field interaction, making it possible to accelerate convergence of scattering observables with respect to the maximum number of rotational states and $J$-blocks included in the basis set. Our method is thus particularly suitable for quantum scattering calculations on atom-molecule (and possibly molecule-molecule) collision systems, where different rotational states are strongly coupled by the anisotropy of the  interaction potential.


As in the case of molecular collisions in magnetic fields \cite{jcp10},  truncation of the asymptotic Hamiltonian matrix leads to the appearance of unphysical Stark shifts. We have analyzed the properties of the unphysical states using a simple 6-state model, which shows that the unphysical states arise due to the electric field-induced coupling between different rotational states in adjacent $J$-blocks. The eigenvectors of the unphysical states are linear combinations of different rotational and $J$-states with field-dependent mixing coefficients. Because of the admixture of higher $J$-states, which do not contribute to low-temperature collision observables due to centrifugal barriers, the unphysical states are expected to play no role in cold atom-molecule collisions. Furthermore, our analytical results suggest that, by neglecting certain coupling matrix elements, it may be possible to completely eliminate the unphysical Stark states from scattering calculations. 

In order to test the performance of our method, we applied it to calculate the cross sections for vibrational and Stark relaxation in He-CaD collisions in the presence of an electric field. The results obtained using the BF approach are in good agreement with benchmark calculations based on the fully uncoupled SF representation. Most notably,  the number of BF channels  required to obtain converged results is smaller by a factor of 1.5 to 4 (depending on $J_\text{max}$) leading to a 5-100 fold gain in computational efficiency (see Table I). These improvements  open up the possibility of carrying out highly efficient quantum scattering calculations of strongly anisotropic atom-molecule collisions in electric fields, which are of great current  interest as potential candidate systems for sympathetic cooling experiments \cite{Li-CaH,N-NH,Rb-NH3} or reactants for electric field-controlled chemical reactions \cite{KRb1,KRb2}.





\acknowledgements
The author is grateful to Rosario Gonz{\'a}lez-F{\'e}rez and Roman Krems for their interest in this work and stimulating discussions.
This work was supported by NSF grants to the Harvard-MIT Center for Ultracold Atoms and the Institute for Theoretical Atomic, Molecular and Optical Physics at Harvard University and the Smithsonian Astrophysical Observatory.

\newpage

Table I. The number of channels in BF basis sets with different $J_\text{max}$   for $M=0$. All the basis sets include 2 vibrational and 10 rotational states of CaD($^2\Sigma^+$) and 10 partial waves (for the SF basis). The ratio $(N_{\text{SF}}/N_\text{BF})^3 $ quantifies the computational efficiency gained by using the BF approach. The number of channels in the fully uncoupled SF representation $N_\text{SF}=1340$.


\vspace{0.6cm}
\begin{center}
\begin{tabular}{ccc}
\hline
\hline
 $J_\text{max}$ &  $\quad$ $N_\text{BF}$ & $\quad$ $(N_{\text{SF}}/N_\text{BF})^3$ \\
\hline
  3 & $\quad$ 280  &  $\quad$  109.6 \\
  4 & $\quad$ 420  &  $\quad$  32.5 \\
   5 & $\quad$ 580  &  $\quad$ 12.3 \\
  6 & $\quad$ 756  &  $\quad$ 5.6 \\
\hline
\hline
\end{tabular}
\end{center}

\newpage

\newpage


\begin{thebibliography}{99}


\bibitem{Book}
{\it  Cold Molecules: Theory, Experiment, Applications}, edited by R. V. Krems, W. C. Stwalley, and B. Friedrich (CRC press, New York, 2009).

\bibitem{NJP}
L. D. Carr, D. DeMille, R. V. Krems, and J. Ye, New J. Phys. {\bf 11}, 055049 (2009).

\bibitem{Schnell}
M. Schnell and G. Meijer, Angew. Chem. Int. Ed. {\bf 48}, 6010 (2009).

\bibitem{Softley}
M. T. Bell and T. P. Softley, Mol. Phys. {\bf 107}, 99 (2009).

\bibitem{Dulieu}
O. Dulieu and C. Gabbanini, Rep. Prog. Phys. {\bf 72}, 086401 (2009).

\bibitem{JohnCPC}
B. Friedrich and J. M. Doyle, ChemPhysChem {\bf 10}, 604 (2009).


\bibitem{Roman08}
R. V. Krems, Phys. Chem. Chem. Phys. {\bf 10}, 4079 (2008).


\bibitem{KRb1}
K.-K. Ni, S. Ospelkaus, D. Wang, G. Qu{\'e}m{\'e}ner, B. Neyenhuis, M. H. G. de Miranda, J. L. Bohn, J. Ye, and D. S. Jin, Nature (London) {\bf 464}, 1324 (2010).

\bibitem{KRb2}
M. H. G. de Miranda, A. Chotia, B. Neyenhuis, D. Wang, G. Qu{\'e}m{\'e}ner, S. Ospelkaus, J. L. Bohn, J. Ye, and D. S. Jin, Nature Phys. {\bf 7}, 502 (2011).

\bibitem{BalaJPB04}
E. Bodo, F. A. Gianturco, N. Balakrishnan, and A. Dalgarno, \jpb{37}{3641}{2004}.

\bibitem{prl06}
T. V. Tscherbul and R. V. Krems, Phys. Rev. Lett. {\bf 97}, 083201 (2006).


\bibitem{jcp06}
T. V. Tscherbul, and R. V. Krems, \jcp{125}{194311}{2006}.

\bibitem{jcp07}
E. Abrahamsson, T. V. Tscherbul, and R. V. Krems, \jcp{127}{044302}{2007}.


\bibitem{Sergey09}
S. V. Alyabyshev, T. V. Tscherbul, and R. V. Krems, \pra{79}{060703(R)}{2009};~S.~V. Alyabyshev and R. V. Krems, \pra{80}{033419}{2009}. 

\bibitem{jcp08}
T. V. Tscherbul and R. V. Krems, \jcp{129}{034112}{2008}.

\bibitem{MeyerBohn}
E. R. Meyer and J. L. Bohn, \pra{82}{042702}{2010}.

\bibitem{KRb_theory}
Z. Idziaszek, G. Qu\'em\'ener, J. L. Bohn, and P. S. Julienne, Phys. Rev. A {\bf 82}, 020703(R) (2010);
G. Qu\'em\'ener, J. L. Bohn, A. Petrov, and S. Kotochigova, Phys. Rev. A {\bf 84}, 062703 (2011);
P. S. Julienne, T. M. Hanna, and Z. Idziaszek, Phys. Chem. Chem. Phys. {\bf 13}, 19114 (2011).



\bibitem{CaH}
J. D. Weinstein, R. deCarvalho, T. Guillet, B. Friedrich, and J. M. Doyle, Nature (London) {\bf 395}, 148 (1998).

\bibitem{He-OH}
B. C. Sawyer, B. K. Stuhl, D. Wang, M. Yeo, and J. Ye, Phys. Rev. Lett. {\bf 101}, 203203 (2008).

\bibitem{RbCs}
E. R. Hudson, N. B. Gilfoy, S. Kotochigova, J. M. Sage, and D. DeMille, \prl{100}{203201}{2008}.

\bibitem{FeshbachChemistry}
S. Knoop, F. Ferlaino, M. Berninger, M. Mark, H.-C. N{\"a}gerl, R. Grimm, J. P. D'Incao, and B. D. Esry, \prl{104}{053201}{2010}.

\bibitem{NHexp}
W. C. Campbell, E. Tsikata, H.-I Lu, L. D. van Buuren, and J. M. Doyle, \prl{98}{213001}{2007}.

\bibitem{NH}
W. C. Campbell, T. V. Tscherbul, H.-I Lu, E. Tsikata, R. V. Krems, and J. M. Doyle, Phys. Rev. Lett. {\bf 102}, 013003 (2009).

\bibitem{N-NH}
M. T. Hummon, T. V. Tscherbul, J. K{\l}os, H.-I Lu, E. Tsikata, W. C. Campbell, A. Dalgarno, and J.~M.~Doyle, \prl{106}{053201}{2011};
P. {\.Z}uchowski and J. M. Hutson, Phys. Chem. Chem. Phys. {\bf 13}, 3669 (2011).


\bibitem{OH-ND3}
B. C. Sawyer, B. K. Stuhl, M. Yeo, T. V. Tscherbul, M. T. Hummon, Y. Xia, J. K{\l}os, D. Patterson, J.~M.~Doyle, and J. Ye,  Phys. Chem. Chem. Phys. {\bf 13}, 19059 (2011).

\bibitem{Rb-NH3}
L. P. Parazzoli, N. J. Fitch, P. S. {\.Z}uchowski, J. M. Hutson, and H. J. Lewandowski, \prl{106}{193201}{2011}.


\bibitem{AD}
A. M. Arthurs and A. Dalgarno, Proc. R. Soc. London Ser. A {\bf 256}, 540 (1960).

\bibitem{Lester}
W. A. Lester, Jr., in {\it Dynamics of Molecular Collisions}, edited by W. H. Miller (Plenum, New York, 1976). 


\bibitem{VolpiBohn}
A. Volpi and J. L. Bohn, Phys. Rev. A {\bf 65}, 052712 (2002).

\bibitem{Roman04}
R. V. Krems and A. Dalgarno, J. Chem. Phys. {\bf 120}, 2296 (2004).

\bibitem{Jeremy07}
M. L. Gonz{\'a}lez-Mart{\'i}nez and J. M. Hutson, \pra{75}{022702}{2007}. 


\bibitem{HeCH2}
T. V. Tscherbul, H.-G. Yu, and A. Dalgarno, \prl{106}{073201}{2011}.

\bibitem{Thierry}
F. Turpin, T. Stoecklin, and P. Halvick, \pra{83}{032717}{2011};  T. Stoecklin and P. Halvick, Phys. Chem. Chem. Phys. {\bf 13}, 19142 (2011).

\bibitem{Chinese}
E. Feng, C. Yu, C. Sun, X. Shao, and W. Huang , \pra{84}{062711}{2011}; 
E. Feng, X. Shao, C. Yu, C. Sun, and W. Huang, \jcp{136}{054302}{2012}.

\bibitem{njp09}
T. V. Tscherbul, Yu. V. Suleimanov, V. Aquilanti, and R. V. Krems, New J. Phys. {\bf 11}, 055021 (2009).

\bibitem{JesusO2}
J. P{\'e}rez-R{\'i}os, J. Campos-Mart{\'i}nez, and M. I. Hern{\'a}ndez, \jcp{134}{124310}{2011}.


\bibitem{Liesbeth}
L. M. C. Janssen, P. S. {\.Z}uchowski, A. van der Avoird, G. C. Groenenboom, and J. M. Hutson, \pra{83}{022713}{2011}.


\bibitem{Li-NH}
A. O. G. Wallis, E. J.J. Longdon, P. S. {\.Z}uchowski, and J. M. Hutson, Eur. Phys. J. D {\bf 65}, 151 (2011).

\bibitem{jcp10}
T. V. Tscherbul and A. Dalgarno, J. Chem. Phys. {\bf 133}, 184104 (2010).

\bibitem{Li-CaH}
T. V. Tscherbul, J. K{\l}os, and A. A. Buchachenko, Phys. Rev. A {\bf 84}, 040701(R) (2011).

\bibitem{Yura}
Y. V. Suleimanov,  T. V. Tscherbul, and R. V. Krems (2012), arXiv:1202.0597 [physics.atom-ph].

\bibitem{AvdeenkovBohn}
A. V. Avdeenkov and J. L. Bohn, Phys. Rev. Lett. {\bf 90}, 043006 (2003).

\bibitem{Ticknor}
C. Ticknor, \prl{100}{133202}{2008}; \pra{76}{052703}{2007};
J. L. Bohn, M. Cavagnero, and C. Ticknor, New J. Phys. {\bf 11}, 055039 (2009); V. Roudnev and M. Cavagnero, \pra{79}{014701}{2009}.

\bibitem{FaradayDiscuss}
T. V. Tscherbul, G. Groenenboom, R. V. Krems, and A. Dalgarno, Faraday Discuss. {\bf 142}, 127 (2009).

\bibitem{BalaHeCaH_paper1}
G. C. Groenenboom and N. Balakrishnan, \jcp{118}{7380}{2003}.


\bibitem{Zare}
R. N. Zare, {\it Angular momentum} (Wiley, NY, 1988).


\bibitem{note}
Since in this work, we are interested in collisions of $^1\Sigma$ molecules, $S=\Sigma=0$, $k=\Omega$, $j=N$, and $k=K_N$. The quantum numbers $\Sigma$, $\Omega$,  $N$ and $K_N$ are defined in Ref. \cite{jcp10}.

\bibitem{Rosario}
R. Gonz{\'a}lez-F{\'e}rez and P. Schmelcher, \pra{69}{023402}{2004}; \pra{71}{033416}{2005}; Europhys. Lett. {\bf 72}, 555 (2005).

\bibitem{David86}
D. E. Manolopoulos, \jcp{85}{6425}{1986}.

\bibitem{BohnChapter}
J. L. Bohn, in {\it  Cold Molecules: Theory, Experiment, Applications}, edited by R. V. Krems, W. C. Stwalley, and B. Friedrich (CRC press, New York, 2009).

\bibitem{ABC}
D. Skouteris, J. F. Castillo, and D. E. Manolopoulos, Comp. Phys. Commun. {\bf 133}, 128 (2000).


\bibitem{Launay}
J. M. Launay, \jpb{10}{3665}{1977}.

\bibitem{BalaHeCaH}
N. Balakrishnan, G. C. Groenenboom, R. V. Krems, and A. Dalgarno, \jcp{118}{7386}{2003}.

\bibitem{ColbertMiller}
D. T. Colbert and W. H. Miller, \jcp{96}{1982}{1992}.

\bibitem{Bala98}
N. Balakrishnan, R. C. Forrey, and A. Dalgarno, \prl{80}{3224}{1998};
R. C. Forrey, N. Balakrishnan, A. Dalgarno, M. R. Haggerty, and E. J. Heller, \prl{82}{2657}{1999}.



































\end{thebibliography}
\end{document}